\documentclass[a4paper,11pt]{article}
\usepackage{graphicx}
\usepackage{amsfonts}
\usepackage{amssymb}
\begin{document}
\begin{titlepage}
\title{Black Hole Entropy Prediction without Immirzi Parameter}

\maketitle
\begin {center}
\author{\large Brian Kong $^1$, Youngsub Yoon $^2$}

\vskip 1cm

{
\it
%\large
%affiliations
$^1$Milton Academy, Milton, MA 02186, USA

$^2$Harvard University, Cambridge, MA 02138, USA
}

\end{center}

\begin{abstract}
In our earlier paper ``Corrections to the Bekenstein-Hawking entropy and the Hawking radiation spectrum,''arXiv:0910.2755, we provided two concrete numerical evidences for the new area spectrum based on the Einstein-Kaufman pseudo tensor as opposed to the Ashtekar variables: namely, the reproduction of the Bekenstein-Hawking entropy without fixing Immirzi parameter and the reproduction of the Hawking radiation spectrum. In this article, we provide another concrete, numerical evidence for this new area spectrum; there was a constant in our earlier article, which was inversely proportional to the density of state, and which we could not fix a priori. Nevertheless, in our earlier article, we obtained this constant to be around 172$\sim$173 by fitting it to the Planck radiation spectrum. In this article, we calculate this value using another method. We obtain 172.87...which implies consistency.

\end{abstract}

\end{titlepage}
\pagebreak

\section {Introduction}
In our earlier paper \cite{corrections} we derived the area spectrum by using the Einstein-Kaufman pseudo tensor, by pointing out an error in the previous derivation using the Ashtekar variables. We showed that not the length of the gravitational electric field based on the Ashtekar variables, but the length of our ``new gravitational electric field'' based on the Einstein-Kaufman pseudo tensor gives the correct area spectrum. To support this proposal, we gave two concrete evidences: the reproduction of the Bekenstein-Hawking entropy without fixing Immirzi parameter and the reproduction of the Hawking radiation spectrum. However, there was a proportionality constant that we couldn't determine in the Hawking radiation spectrum. In other words, we reproduced the Hawking radiation spectrum only up to relative intensity. Nevertheless, we obtained the proportionality constant by numerical fitting. In this article, we analytically calculate this value, and confirm that it indeed agrees with the one obtained by fitting.

To this end, we will closely follow \cite{Bekenstein}. In that article, Bekenstein and Mukhanov calculated the black hole radiation spectrum assuming that the area is quantized in integer multiples of a unit area, which we now know that it's not the case. However, much of their arguments are still valid, if one replaces the equations appropriately. Closely following their derivations enables us to calculate the proportionality constant.

\section {Spectroscopy of the quantum black hole by Bekenstein and Mukhanov}

From the same procedure as in \cite{Bekenstein}, we can obtain the following formula.

\begin{equation}
x_a=\frac{\Delta t}{\tau}\frac{f(A_{BH}-a)}{f(A_{BH})} \label{23}
\end{equation}

Where $A_{BH}$ is the area of the black hole, and $f(A)$ is the degeneracy of the black hole with area $A$, and $a$ is the area spectrum and $x_a$ is the average number of the photons, of which the frequency corresponds to the area spectrum $a$, emitted during time $\Delta t$, and $\tau$ is a time scale that is given by the following formula.

\begin{equation}
\Delta M=-\frac{\int dn_{photon}h\nu }{\int dn_{photon}}\frac{\Delta t}{\tau}\label{19}
\end{equation}

This formula fixes $\tau$. Here, $dn_{photon}$, the number of photons with given frequency between $\nu$ and $\nu+d\nu$ emitted during unit time, is given by the following formula:

\begin{equation}
dn_{photon}=\frac{2\nu^2}{c^2}\frac{A_{BH}d\nu}{e^{h\nu/kT}-1}\label{dnphoton}
\end{equation}
What (\ref{19}) tells us is that the energy lost by the Hawking radiation during time $\Delta t$ is given by the average of the energy of photon emitted during this time multiplied by $1/\tau$. Using the following condition:

\begin{equation}
\frac{\Delta M}{\Delta t}=-\int dn_{photon}h\nu
\end{equation}
which is obvious since $h\nu$ is the energy of a single photon with the frequency $\nu$, we obtain the following:
\begin{equation}
\frac{1}{\tau}=\int dn_{photon}\label{tau}
\end{equation}

Also, the degeneracy function $f$ in (\ref{23}) must satisfy the following condition by consistency argument in \cite{Bekenstein}:
\begin{equation}
1=\sum_a \frac{f(A_{BH}-a)}{f(A_{BH})} \label{surprisingly}
\end{equation}

Surprisingly, this equation was automatically satisfied in \cite{corrections}. This is actually how we showed that our new area spectrum reproduces the correct black hole entropy.

\section {Tying the number of degeneracy into the Hawking radiation spectrum}
Given this, how can we relate (\ref{23}) with (\ref{dnphoton})?

Using the notation of \cite{corrections}, we can relate them by the following formula:
\begin{equation}
x_a (N(a+\Delta a)-N(a))=\Delta t \frac{2\nu^2}{c^2}\frac{A_{BH}\Delta\nu}{e^{h\nu/kT}-1}\label{tying}
\end{equation}
where $N(a+\Delta a)-N(a)$ is the number of states between $a$ and $a+\Delta a$. Here, $a$ and $\nu$ are related by the following formula:
\begin{equation}
\frac{a}{4}=\frac{h\nu}{kT}
\end{equation}
because of the following condition:

\begin{equation}
\frac{1}{e^{a/4}-1}=\frac{1}{e^{h\nu/kT}-1}
\end{equation}

Again, using the notation of \cite{corrections}, we have the following definition for $C$:
\begin{equation}
N(a+\Delta a)-N(a)=\frac{I(a+\Delta a)-I(a)}{C}=\frac{\sqrt{a}(e^{a/4}-1)\Delta a}{C}\label{N}
\end{equation}

Now, we will explain what a natural and convenient choice for $\Delta a$ would be:

From (41) in \cite{corrections}, for every area spectrum $a$ we can obtain an integer $y(a)$ such that the following is satisfied:
\begin{equation}
a=4\sqrt{2}\pi y(a)^{1/4}
\end{equation}

Then, we can have the following:
\begin{equation}
a+\Delta a=4\sqrt{2}\pi (y(a)+1)^{1/4}
\end{equation}

Therefore, for large $a$, we get:
\begin{equation}
\Delta a=\frac{256\pi^4}{a^3}\label{Deltaa}
\end{equation}

So far we have focused more or less on the left-hand-side of (\ref{tying}). We can easily substitute (\ref{tau}), (\ref{N}) and (\ref{Deltaa}) to the left-hand-side. Now, let's focus on the right-hand-side. To this end, let's digress to the topic of density of state. For a cube with size $L\times L\times L$, we have the following formulas for the momentum:

\begin{eqnarray}
p_x=\frac{h n_x}{2L}\nonumber\\
p_y=\frac{h n_y}{2L}\nonumber\\
p_z=\frac{h n_z}{2L}
\end{eqnarray}

By relating the momentum and the frequency of photons, and by obtaining the density of state using standard procedure, we get the following formula:

\begin{equation}
\frac{h^3}{8L^3}(n^2 \Delta n)=p^2\Delta p=\frac{h^3}{c^3}\nu^2 \Delta\nu
\end{equation}

A natural choice for $\Delta n$ that corresponds to the choice for the left-hand-side of (\ref{tying}) is following:
\begin{equation}
2(\frac{\pi}{2}n^2 \Delta n)=1
\end{equation}
where $\frac{\pi}{2} n^2 dn$ comes from the surface area of one-eighth of a sphere, as $n_x$, $n_y$, $n_z$ are positive and the factor 2 comes from the two polarizations of photons.

Therefore the right-hand-side of (\ref{tying}) becomes:
\begin{equation}
\frac{c}{4\pi L^3}\frac{A_{BH}}{e^{a/4}-1} \Delta t\label{righthandside}
\end{equation}

As $L^3$ is the volume of the cube, we can regard it as the volume of black hole, even though the shape of the black hole is not rectangular, but sphere; we would have obtained the same result, if we considered the density of state of photons confined in sphere, even though the actual calculation would be different and more complicated. All that matters is that we recover the volume factor for $L^3$. Given this, using the following relations,
\begin{eqnarray}
A_{BH}=4\pi r^2\nonumber\\
L^3=\frac{4\pi r^3}{3}
\end{eqnarray}
we get:
\begin{equation}
L^3=\frac{A_{BH}\sqrt{A_{BH}}}{6\sqrt{\pi}}\label{Lcube}
\end{equation}

Now, we can plug in (\ref{Lcube}) to (\ref{righthandside}), and then to (\ref{tying}). However, we must be careful when we calculate $\tau$. As was shown in \cite{corrections}, there is no light emitted below certain frequency in the Hawking radiation. Considering this, $\tau$ is given by the following expression:

\begin{equation}
\frac{1}{\tau}=A_{BH}\frac{2}{c^2}\int_{\pi\sqrt{2}}^{\infty}\frac{u^2 du}{e^u-1}(\frac{kT}{h})^3
\end{equation}

Using the fact that $kT=1/(8\pi M)$ and $A_{BH}=16\pi M^2$, and tying everything together, we obtain:
\begin{equation}
\frac{f(A_{BH}-a)}{f(A)}=\frac{3C}{128\pi^3\alpha}\frac{a^{5/2}}{(e^{a/4}-1)^2}\label{ff}
\end{equation}
where $\alpha$ is given by:
\begin{equation}
\int_{\pi\sqrt{2}}^{\infty}\frac{u^2 du}{e^u-1}=0.36193...
\end{equation}
Recalling (\ref{surprisingly}), we can take the summation with respect to $a$ in the both-hand-sides of (\ref{ff}). This yields the following:

\begin{equation}
1=\frac{3C}{128\pi^3\alpha}\sum_a \frac{a^{5/2}}{(e^{a/4}-1)^2}
\end{equation}

Using the following calculation,
\begin{equation}
\sum_a \frac{a^{5/2}}{(e^{a/4}-1)^2}=2.7697...
\end{equation}

we obtain:

\begin{equation}
C=172.87...
\end{equation}
which agrees with the value 172$\sim$173 which we obtained in \cite{corrections}.

\section {Discussion and Conclusions}
In this paper, we provided the third evidence for the area spectrum based on the Einstein-Kaufman pseudo tensor. We recovered the proportionality constant which used to be known only through fitting. That this value $C$ agrees with the value obtained in \cite{corrections} gives another strong support on our area spectrum based on the Einstein-Kaufman pseudo tensor. One may also check whether this consistency that the two different methods yield the same value of $C$ hold in the case of the Ashtekar variable theory with the corresponding $C$ defined appropriately. Depending on the result, this may either falsify or support the area spectrum based on the Ashtekar variable.

\pagebreak

\end{document}